\documentclass[prd,
twocolumn,
superscriptaddress,nofootinbib,%
tightenlines,shownopacs]{revtex4}

\usepackage{graphicx}  % needed for figures
\usepackage{dcolumn}   % needed for some tables
\usepackage{bm}        % for math
\usepackage{amssymb}   % for math
\usepackage{slashed}
\usepackage{feynmf}
\usepackage{tensor}
\usepackage{amsmath,amssymb,amsfonts,epsfig,graphicx,euscript}
\usepackage{graphics}
\newcommand{\bse}{\begin{subequations}}
\newcommand{\ese}{\end{subequations}}
\newcommand{\be}{\begin{equation}}
\newcommand{\ee}{\end{equation}}
\newcommand{\bea}{\begin{eqnarray}}
\newcommand{\eea}{\end{eqnarray}}
\newcommand{\ba}{\begin{array}}
\newcommand{\ea}{\end{array}}

\begin{document}

\title{Non-Relativistic Fermion-Fermion Scattering in Higher Derivative Gravity}

\author{Mohammad A. Ganjali}
\email{ganjali@theory.ipm.ac.ir}
\author{Vahid Amirkhani}
\email{std_ amirkhani@khu.ac.ir}
\affiliation{Department of Physics, Kharazmi University, P. O. Box 31979-37551, Tehran, Iran}
\author{Ahmad ShamlouMehr}%
\email{as876016@ohio.edu}
\affiliation{Department of Physics and Astronomy, Ohio University, 251B Clippinger Labs, Athens, OH 45701, USA}

\begin{abstract}
In this note, we examine the scattering of two identical fermions in theories where fermionic fields minimally coupled to higher derivative gravity. In particular, we consider the extension of general relativity with  $R^2$ corrections or non-local terms. We expand the action of fermions around the flat space background and obtain two fermion-one graviton vertex. Then, by considering the scattering amplitude of two fermions, we calculate the non-relativistic limit and that obtain the potential for two fermion-fermion interaction which would be the usual Newtonian potential corrected with a Yukawa-like term. At the end, we briefly discuss the astronomical effects of such Yukawa-like potential by computing the gravitational pressure of a spherical star and use it for a white dwarf to obtain quantum corrections of Chandrasekhar radius.
\end{abstract}

%\pacs{}
\maketitle

\section{Introduction}
Extended theories of gravity are extensions of Einstein theory of gravity in which include
higher power of the Riemann, Ricci, Weyl tensors, curvature scalars or their derivatives \cite{Stelle:1976gc}. Having several interesting properties causes that lots of attention recently attracted to these higher derivative theories and their various aspects such as their classical solutions, quantization, unitarity, renormalizability, applications in particle physics, cosmology and etc, have been studied and some very interesting and remarkable results have been found \cite{Boulware:1983td}, see \cite{Bueno:2016ypa} for a recent good review on aspects of higher derivative gravity. Quantization of these extended gravity is one of the most important subjects in this line and various important results have obtained. In more detail, these theories suffer from the unitarity and renormalizability.

In a recent interesting paper \cite{Alvarez-Gaume:2015rwa}, the author argued that the physical propagating modes of $R^2$ theory depends on the specific background. In particular, for flat backgrounds the pure $R^2$ theory propagates only a scalar massless mode. Propagation of a spin-two field in this theory occurs when one expands the theory around the curved backgrounds(like de Sitter or anti-de Sitter) or by adding to theory the standard Einstein term. In the latter case, one obtains an $R^2$ theory which is relevant for short distance physics in UV and gravitate in the IR which is responsible for the long range gravity. Adding the Einstein term causes that a massless zero spin appears and would give a mass and a massless zero spin ghost also enters the game but, it was shown that such spin zero ghost field is harmless and non-physical.
One may also add the Weyl$^2$ term to the action and study the spectrum of this theory. The action for this theory can be written as:
\begin{equation}
\label{action}
S=\int{d^4x\sqrt{-g}\left(a(R_{\mu\nu}R^{\mu\nu}-\frac{1}{3}R^2)+bR^2-\kappa^2 R+\lambda\right)}.
\end{equation}
However, in this theory, besides the propagation of a massive scalar with mass $m^2=-\kappa^2 /6b$ and a massless tensor fields, there are also nonphysical massless spin zero ghost and the usual massive spin 2 ghost with the mass $\kappa^2/a$ for\footnote{When $a>0$ this state would be tachyonic.} $a<0$ \cite{Julve:1978xn}.

Beside of this higher curvature theory, one may add terms with higher derivatives of scalar tensors and so on. In particular, the following action has been proposed in \cite{Maggiore:2013mea}
\begin{equation}
\label{actionnon}
S=\kappa^2\int{d^4x\sqrt{-g}\left(R-\frac{1}{6}m'^2R\frac{1}{\square^{2}}R\right)},
\end{equation}
where $\square$ is d'Alembertian operator in four dimensional flat space and $m'$ is a parameter with dimension \textit{mass}.

Higher derivatives of fields in the action usually make the theory nonlocal. It was also shown that the spectrum of the theory includes a massless graviton, a massless scalar and a ghostlike massive scalar. But, as it was argued in \cite{Maggiore:2013mea}, both  non-locality and presence of ghostlike field come from the fact that the action (\ref{actionnon}) is at the level of an effective action and that at a fundamental level such problems would be disappear.

Many attempts have been made to study these theories especially, their quantum field theoretical aspects and their applications in cosmology and a possible quantum gravitational extension of standard model of particle physics.

In this note, we will consider the actions (\ref{action}) and (\ref{actionnon}) where will be minimally coupled to fermions and by using the propagators of scalar and tensor modes, we briefly study fermion-fermion scattering in these gravitational theories. In particular, our aim is to compute the corrected Newtonian potential for two fermion interaction at the above $R^2$ and non-local gravity theory.

The paper organized as follows: In section 2 we will find Lagrangian for fermion-graviton interaction. In section 3 we calculate the amplitude of electron-electron scattering due to graviton exchange for higher derivative theories which is introduced in section 1 and discuss the Newtonian limit of them.  In section 4 we investigate astronomical effects of Yukawa-like term which is appeared in Newtonian potential. Finally, the main points are summarized in section 5.

\section{Electron-Graviton Vertex}
In this section, we consider the minimal coupling of fermionic fields with gravity in higher derivative theories with gravitational actions (\ref{action}) or (\ref{actionnon}) by adding the following action:
%\bea\label{faction}
%S_f=\int{d^4x\sqrt{-g}\left(\frac{i}{2}\bar{\psi}\gamma^{\mu}\partial_{\mu}\psi
%-\frac{i}{2}\partial_{\mu}\bar{\psi}\gamma^{\mu}\psi+
%\frac{i}{2}\bar{\psi}\{\gamma^{\mu},\Omega_{\mu}\}\psi-
%m\bar{\psi}\psi\right)}.
%\eea
\begin{equation}
\label{faction}
S_f=\int{d^4x\sqrt{-g}\left(\frac{i}{2}(\bar{\psi}\gamma^{\mu}D_{\mu}\psi
-D_{\mu}\bar{\psi}\gamma^{\mu}\psi)-
m\bar{\psi}\psi\right)},
\end{equation}
where we have defined $D_{\mu}=\partial_{\mu}-\Omega_{\mu}$ and
\begin{equation}
\Omega_{\mu}=
\frac{i}{2}\eta_{ac}e^a_{\;\nu}\left(\partial_{\mu}e^{\nu}_{\;b} + e^{\sigma}_{\;b}\Gamma^{\nu}_{\mu\sigma}\right), \quad \gamma^{\mu}=\gamma^ae^{\mu}_{\;a},
\end{equation}
We note that $\gamma^a$'s are the usual Dirac gamma matrices in four dimensions and $e^{\mu}_a$ are the veilbein of the given geometry and satisfy $g_{\mu\nu}e^{\mu}_{a}e^{\nu}_{b}=\eta_{ab}=diag(+,-,-,-)$.

Now, we would like to find the Lagrangian for electron-graviton interaction. As usual, one should expand the above action around a background geometry and reads the corresponding interaction terms. Expansion up to linear order would give us the tree level vertexes. In this paper, we will expand the metric around the flat space geometry as $g_{\mu\nu}=\eta_{\mu\nu}+h_{\mu\nu}$.

Expanding the various terms of the action Eq.~(\ref{faction}) is straightforward but one should care about the expansion of the veilbeins as follows. We suppose that $e^{\mu}_{\;a}=\delta^{\mu}_{\;a}+\chi^{\mu}_{\;a}$ where $\chi=ch+{\cal O}(h^2)$ and $c$ is a constant and should be determined such that $g_{\mu\nu}e^{\mu}_ae^{\nu}_b=\eta_{ab}$. Furthermore, we have a freedom in fixing these $e^{\mu}_{\;a}$'s due to the local $GL(4,R)$ symmetry acting on veilbeis in the action Eq.~(\ref{faction}). Using this symmetry, one may consider $\eta_{\mu\nu}\delta^{\mu}_{\;a}\chi^{\nu}_{\;b}=\eta_{\mu\nu}\delta^{\nu}_{\;b}e^{\mu}_{\;a}$ and that obtains:
\begin{align}
\begin{split}\label{veilbein}
e^{\mu}_{\;a}&=\delta^{\mu}_{\;a}-\frac{1}{2}\eta^{\lambda\mu}h_{\lambda\nu}\delta^{\nu}_{\;a}+{\cal O}(h^2) \\
e^{a}_{\;\mu}&=\delta^{a}_{\;\mu}+\frac{1}{2}\eta^{\nu\lambda}h_{\lambda\mu}\delta^{a}_{\;\mu}+{\cal O}(h^2).
\end{split}
\end{align}
Using (\ref{veilbein}) and expanding (\ref{faction}) around the flat space metric and performing a bit of calculation and then applying the integration by part, one obtains the following Lagrangian for fermion-graviton interaction
\bea
{\cal L}_{fg}=\tau^{\mu\nu}h_{\mu\nu},
\eea
where
\begin{widetext}
\begin{align}
\begin{split}
\label{fgaction}
\tau^{\mu\nu}=&\;\frac{i}{2}\bar{\psi}\gamma^a\left(\eta^{\mu\nu}\delta_{\;a}^{\alpha}
-\eta^{\mu\alpha}\delta_{\;a}^{\nu}\right)\partial_{\alpha}\psi-
\frac{i}{2}\partial_{\alpha}\bar{\psi}\gamma^a\left(\eta^{\mu\nu}\delta_{\;a}^{\alpha}
-\eta^{\mu\alpha}\delta_{\;a}^{\nu}\right)\psi \cr
& +\frac{1}{4}\partial_{\alpha}\left(\bar{\psi}\{\gamma^a\delta^{\mu}_{\;a},\eta^{bc}
\delta^{a}_{\;\theta}\delta^{\alpha}_{\;c}\eta^{\theta\nu}\Sigma_{ab}\}\psi\right)-
\frac{1}{4}\partial_{\alpha}\left(\bar{\psi}\{\gamma^a\delta^{\mu}_{\;a},\eta^{bc}
\delta^{a}_{\;\theta}\delta^{\nu}_{\;c}\eta^{\theta\alpha}\Sigma_{ab}\}\psi\right)%\cr
-\frac{1}{2}m\bar{\psi}\eta^{\mu\nu}\psi,
\end{split}
\end{align}
\end{widetext}
where it was defined $\Sigma^{ab}=\frac{i}{4}\left[\gamma^a,\gamma^b\right]$.

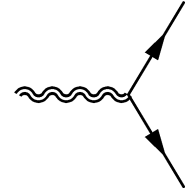
\begin{figure}[h]
\centering
\begin{fmfgraph*}(70,70)
    \fmfleft{i1}
    \fmfright{o1,o2}
    \fmf{dbl_wiggly}{i1,w1}
    \fmf{fermion,label=$p_{i}$}{w1,o2}
    \fmf{fermion,label=$p_{j}$}{w1,o1}
    %\fmfv{lab=$V^{\ast}_{ud}$,lab.dist=0.05w}{w1}
\end{fmfgraph*}
%end{fmffile}
\medskip
\caption{\label{fig:vertex}electron-graviton vertex}
\end{figure}

Note that up to now, we restrict ourself to distinguish between the coordinate frame and local frame. In the next sections, it can be seen that when we obtain the vertexes and do subsequent computations, we can use all local frame relations such as the usual flat space gamma matric relations and etc. Thus, hereafter, we will use the greek indices as local frame notations! With this consideration, it is not hard to see that the two electrons-one graviton vertex with ingoing and outgoing fermions with momenta $p_i$ and $p_j$, Fig.~\ref{fig:vertex}, can be written as
\bea\label{vertex}
\tilde{\tau}^{\mu\nu}(p_i,p_j)=-ip_{i\alpha}\Gamma^{\mu\nu\alpha}+ip_{j\alpha}\tilde{\Gamma}^{\mu\nu\alpha}
-\frac{1}{2}m\eta^{\mu\nu},
\eea
where we have defined $\Gamma^{\mu\nu\alpha}=\frac{1}{2}\gamma^{\mu}\Sigma^{\nu\alpha}$ and $\tilde{\Gamma}^{\mu\nu\alpha}=\frac{1}{2}\Sigma^{\nu\alpha}\gamma^{\mu}$. Note also that the above vertex is for the case where the initial and final fermions have the equal mass $m$.

\begin{fmffile}{diagram}

\section{$\mathbf{e^-e^-\rightarrow e^-e^-}$ Due to Gravity}
Now, we would like to study the $e^-e^-\rightarrow e^-e^-$ scattering due to the graviton exchange at tree level.
\\[0.5cm]
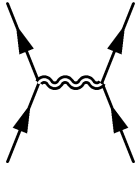
\begin{figure}[h]
\centering
%\begin{fmffile}{diagram}
\begin{fmfgraph*}(60,60)
           \fmfleft{i1,i2}
           \fmflabel{$p_{1}$}{i1}
          \fmflabel{$p_{3}$}{i2}
           \fmfright{o1,o2}
           \fmflabel{$p_{2}$}{o1}
           \fmflabel{$p_{4}$}{o2}
          \fmf{fermion}{i1,v1,i2}
           \fmf{fermion}{o1,v3,o2}
           \fmf{dbl_wiggly, label=$q$}{v1,v3}
            %\fmf{dbl_wiggly}{v2,v3}
       \end{fmfgraph*}
\medskip
\caption{\label{fig:scattering}electron-electron scatterng due to graviton exchange}
\end{figure}
\end{fmffile}

For this purpose, the matrix element is given by \cite{Donoghue:1994dn}
\bea
{\cal M}=\tilde{\tau}(p_1,p_3)\Delta_{\mu\nu,\rho\sigma}\tilde{\tau}(p_2,p_4),
\eea
where $\Delta_{\mu\nu,\rho\sigma}$ includes the propagators for the zero spin scalar field, spin two graviton and spin two ghost field. Using the following projectors:
\bea\label{project}
\begin{split}
P_{\mu\nu,\rho\sigma}^{(2)}&=&\frac{1}{2}\left(
\theta_{\mu\nu}\theta_{\rho\sigma}+\theta_{\mu\rho}\theta_{\nu\sigma}\right)
-\frac{\zeta}{3}\theta_{\mu\nu}\theta_{\rho\sigma},\cr
P_{\mu\nu,\rho\sigma}^{(0)}&=&\frac{1}{3}\theta_{\mu\nu}\theta_{\rho\sigma},
\;\;\;\;\;\;\theta_{\mu\nu}=\eta_{\mu\nu}-\frac{q_{\mu}q_{\nu}}{q^2},
\end{split}
\eea
where $P_{\mu\nu,\rho\sigma}^{(2)}$, $P_{\mu\nu,\rho\sigma}^{(0)}$ are the spin-2 and spin-0 part of $h_{\mu\nu}$ in momentum space and $\zeta$ is a constant number which typically is different for various gravitational theories and reflects the vDVZ discontinuity then, the explicit form of the propagator can be written as: \cite{Stelle:1976gc,Alvarez-Gaume:2015rwa,Maggiore:2013mea,Modesto:2017hzl}
\begin{equation}\label{propagator}
%\begin{split}
\Delta_{\mu\nu,\rho\sigma}^{(2)}=\zeta^{(2)}P_{\mu\nu,\rho\sigma}^{(2)},\;\;\;\;\;\;\;\;
\Delta_{\mu\nu,\rho\sigma}^{(0)}=\zeta^{(0)}P_{\mu\nu,\rho\sigma}^{(0)},
%\end{split}
\end{equation}
and
\begin{equation}
\Delta_{\mu\nu,\rho\sigma}=\Delta_{\mu\nu,\rho\sigma}^{(2)}+\Delta_{\mu\nu,\rho\sigma}^{(0)},
\end{equation}
where for $R^2$ theory one has 
\begin{gather}
\label{R}
\begin{split}
\zeta =1,\;\;\;\;
\zeta^{(2)}=-\frac{2}{q^2(aq^2-\kappa^2)}, \\
\zeta^{(0)}=-\frac{1}{6b}\frac{1}{q^2(q^2+\kappa^2/6b)},
\end{split}
\end{gather}
and for nonlocal theory 
\begin{equation}
\label{non}
\zeta=\frac{3}{2},\;\;\;\;\zeta^{(2)}=\frac{1}{q^2},\;\;\;\;\zeta^{(0)}=\frac{-1}{6}\frac{m'^{2}}{q^2(q^2-m'^2)}.
\end{equation}

Now, using the fact that for the diagram (Fig.~\ref{fig:scattering}), where the initial and final particles have the same mass, one has $\bar{u}(p_{3})\slashed{q}u(p_{1})=\bar{u}(p_{4})\slashed{q}u(p_{2})=0$ then one can obtain the amplitude as:
\begin{widetext}
\begin{align}
\label{amp}
{\cal M}= -\frac{\zeta^{(2)}}{16}\bigg [& -4 \frac{\zeta}{3} m^{2}  \overline{u}(p_{3}) u(p_{1}) \overline{u}(p_{4})u(p_{2})+ (p_{1}.p_{2}+p_{1}.p_{3})\overline{u}(p_{3}) \gamma^{\nu}u(p_{1}) \overline{u}(p_{4})\gamma_{\nu} u(p_{2}) \cr
&+ \frac{1}{2}(p_{1\sigma}+p_{3\sigma})(p_{2\nu}+p_{4\nu}) \overline{u}(p_{3}) \gamma^{\nu}u(p_{1}) \overline{u}(p_{4})\gamma_{\sigma} u(p_{2}) \bigg]
-\frac{\zeta^{(0)}}{12}m^2\left[\bar{u}(p_3)u(p_1)\bar{u}(p_4)u(p_2)\right].
\end{align}
\end{widetext}
\subsection{Newtonian Limit}
Now, we want to calculate nonrelativistic limit of the amplitude (\ref{amp}). In nonrelativistic limit we have
$(q'-q)^{2} \simeq - \vert \mathbf{q'} - \mathbf{q} \vert ^{2} $ and
$u^{s}(p)=\sqrt{m} \left(\begin{smallmatrix}\xi^{s} \\ \xi^{s}\end{smallmatrix}\right)$
where $\xi^{s}$ is a two-component constant spinor normalized to ${\xi^{s'}}^{\dagger}\xi^{s} = \delta^{s's}$. Then, spinor products are $
\overline{u}^{s'}(p')u^{s}(p) = 2m {\xi^{s'}}^{\dagger}\xi^{s} = 2m \delta^{s's}.
$
Contribution from the terms where include gamma matrices in the nonrelativistic limit are
$
\overline{u}(p')\gamma^{\mu} u(p) =\overline{u}(p')\gamma^{0} u(p) =u^{\dagger}(p')u(p) \approx   2m {\xi '}^{\dagger}\xi,
$
the other terms, $\overline{u}(p')\gamma^{i} u(p)$ can be neglected compared to $\overline{u}(p')\gamma^{0} u(p)$. The factors of $2m$ comes from relativistic normalozation conventions and must be dropped \cite{peskin}.  Therefore, the Newtonian limit of scattering of two identical particles can be written as:
\begin{equation}
\label{1}
\mathcal{M}_{nr} =   \frac{\zeta^{(2)}m^{2}}{4} \left( \frac{\zeta}{3}-1 \right) - \frac{\zeta^{(0)}m^{2}}{12} .
\end{equation}
In order to find Newtonian potential of nonrelativistic amplitude for electron-electron scattering due to graviton exchange we use
\begin{equation}
\label{potential}
V = - \frac{\chi}{(2\pi)^{3}} \int \mathcal{M}_{nr} e^{-i\mathbf{q}.\mathbf{r}} d^{3}q ,
\end{equation}
where for $R+R^{2}+$Weyl$^{2}$  gravity $\chi=1$, but in nonlocal theory $\chi =3/(2\kappa^{2})$ and $\kappa^{2}=1/(16\pi G)$. Note that the coefficient $\chi =3/(2\kappa^{2})$ in front of (\ref{potential}) is necessary to obtain the usual $\frac{1}{r}$ potential from the nonrelativistic limit of non-local gravity (\ref{actionnon}). By using (\ref{1}) and (\ref{potential}) we can calculate Newtonian potential for both theory as follows:
\begin{itemize}
\item \textbf{Newtonian potential for} $\mathbf{R + }\mathbf{R^{2}}+$ \textbf{Weyl}$^{2}$ \textbf{gravity:} In this case, we use (\ref{R})  and (\ref{1}), therefore we obtain: 
\begin{equation}
\mathcal{M}_{nr} =\frac{m^{2}}{\kappa^{2}}\left( \frac{1}{4\vert \mathbf{q}^{2}\vert} - \frac{a}{3 (a \vert \mathbf{q}^{2}\vert + \kappa^{2})} + \frac{b}{2 (6b \vert \mathbf{q}^{2}\vert - \kappa^{2})} \right),
\end{equation}
and the Newtonian potential becomes
\begin{equation}
V = -  \frac{Gm^{2}}{r} + \frac{4Gm^{2}}{3r} e^{-\kappa r/\sqrt{a}} - \frac{Gm^{2}}{3r} e^{-\kappa r/(2\sqrt{3b})}.
\end{equation}
As it was discussed in \cite{Alvarez-Gaume:2015rwa} for 
\begin{equation}
r\gg r_0= \frac{1}{min\left(\frac{1}{\sqrt{2a}}M_P,\frac{1}{2\sqrt{3b}}M_P\right)},
\end{equation}
one obtains the usual Newtonian potential and for $r\ll r_0$ we have a linear potential which is finite at $r=0$ and is confining for $\frac{1}{12b} \preceq \frac{1}{2a}$.

\item \textbf{Newtonian potential for nonlocal gravity:} In this theory, we put (\ref{non}) to (\ref{1}), so nonrelativistic amplitude becomes:
\begin{equation}
\mathcal{M}_{nr} =\frac{m^{2}}{6 \vert \mathbf{q}^{2}\vert } - \frac{m^{2}}{24 \left(\vert\mathbf{q}^{2}\vert + m'^{2} \right) },
\end{equation}
therefore the Newtonian potential is given by 
\begin{equation}
V = -  \frac{Gm^{2}}{r} + \frac{Gm^{2}}{4r} e^{-m'r}.
\end{equation}
\end{itemize}

%%%%%%%%%%%%%%%%%%%%%%%%%%%%%%%%%%%%%%%
\section{Yukawa in Star}
In this section, we search that whether a Yukawa-like term may has any significant effect at astronomical level. In particular, we consider a white dwarf where the gravitational pressure is equal to electron-degeneracy pressure. Thus, we compute the pressure of a homogenous spherical star due to gravity with potential $V(r)=-G\frac{M(r)}{r}+\lambda G\frac{M(r)}{r}e^{-\alpha r}$ which is given by
\begin{equation}
P_g(r)=G\int_r^{R_0}{\frac{M(r)\rho(r)}{r^2}\left(1-\lambda e^{-\alpha r}-\lambda\alpha r e^{-\alpha r}\right)dr},
\end{equation}
where here $\lambda$ and $\alpha$ are some constants, $M(r)$ is the mass of the star up to radius $r$ and $\rho(r)$ and $R_0$ are the mass density and radius of the star respectively. The total mass of the star is $M_0$. Then, with a linear density $\rho(r)=\rho_c(1-\beta\frac{r}{R_0})$ for $r\leq R_0$, where $\rho_c$ and $\beta$ are constants, one straightforwardly obtains
\begin{widetext}
\begin{align}
\label{p}
\begin{split}
P_g(r) =\frac{GM_0^2}{\frac{4}{3}\pi(1-\frac{3}{4}\beta)^2}\frac{\alpha^4}{R^6}
 & \Bigg\{\Big(\frac{1}{2}-\frac{7\beta}{12}
+\frac{3\beta^2}{16}\Big)R^2
-\lambda\Big(3-\frac{14\beta}{R}+\frac{45\beta^2}{2R^2}\Big) - \lambda e^{-R}\bigg [\Big(-1+\frac{7\beta}{4}-\frac{3\beta^2}{4}\Big)R^2 \cr  & +\Big(-3+7\beta-\frac{15\beta^2}{4}\Big)R
+\Big(-3+14\beta-\frac{45\beta^2}{4}\Big)
+\Big(14\beta-\frac{45\beta^2}{2}\Big)\frac{1}{R}+\Big(-\frac{45\beta^2}{2}\Big)\frac{1}{R^2}
\bigg ]\Bigg\},
\end{split}
\end{align}
\end{widetext}
where in (\ref{p}) it was defined $R=\alpha R_0$.
The central electron-degeneracy pressure for non-relativistic particles \cite{chandra} is equal to $(P_e)_c=0.000485\frac{4\pi^2}{m_em_p^{5/3}}(\frac{Z}{A})^{5/3}\rho_c^{5/3}$ where $m_e$ and $m_p$ are the mass of electrons and protons and $Z$ and $A$ are the atomic number and atomic weight.

By equating $(P_g)_c$ with $(P_e)_c$ for the case where $\lambda=0$ one finds the well known Chandrasekhar radius of a white dwarf $R_0^{Ch}=0.00114\frac{4\pi^2}{Gm_em_p^{5/3}}(\frac{Z}{A})^{5/3}M_0^{-1/3}$.
For the nonzero $\lambda$ equating (\ref{p}) with $(P_e)_c$ gives us the Chandrasekhar radius with quantum correction. However, finding the explicit solution from the equation $(P_g)_c=(P_e)_c$ is a very hard task but we can find the equation up to leading order. In fact, in $R^2$ theory for a white dwarf where $R_0\sim 10^7m$ we have $R=\alpha R_0=\sqrt{\frac{c^3}{a\hbar G}}R_0\sim 10^{42}a^{-1/2}$ (or $R=\sqrt{\frac{c^3}{b\hbar G}}R_0\sim 10^{42}b^{-1/2}$). So, for typical values for $a$ or $b$ the amount of $R$ is very large. Thus, by defining $\omega=\frac{1}{2}-\frac{7\beta}{12}+\frac{3\beta^2}{16}$, we obtains the leading order equation as:
\begin{equation}
R-3\frac{\lambda}{\omega}\frac{1}{R}-R^{Ch}=0.
\end{equation}
After all, one easily finds the correction to white dwarf radius due to gravitational quantum effects is of order
\bea
\Delta R\sim \frac{1}{\alpha^2R_0^{Ch}}=\frac{\hbar G}{c^3R_0^{Ch}}.
\eea
In $R^2$ theory, we have two Yukawa-like term in which  $\lambda_1=\frac{4}{3}, \alpha_1=\frac{\kappa}{\sqrt{2a}}$ and $\lambda_2=-\frac{1}{3}, \alpha_2=\frac{\kappa}{2\sqrt{3b}}$. Then $\Delta R_0^{Ch}\sim (8/3 a-4b) \frac{1}{\omega\kappa^2R_0^{Ch}}=(8/3 a-4b)\frac{\hbar G}{c^3R_0^{Ch}}$. Such quantum correction is too small to detect from the astrophysical observations unless the coefficients $a$ or $b$ would be very large. But with larger $a$ or $b$, one should consider higher order corrections to computations of scattering amplitudes and so on. Note also thet for $a=3/2 b$ this quantum correctionis equal to zero.

Similar argument is also true for non-local theory by renaming $m'\rightarrow m'/\kappa$.
%%%%%%%%%%%%%%%%%%%%%%%%%%%%%%%%%%%%%%%%%%%%%%%%%%%%%%%%%%%%%%%%%%%%%%%%%%%%%%%%%%%%%%%%%%%%%%%%%%%%%%%%
\section{Conclusion}
In this paper, we consider the scattering of two fermions in theories where fermionic fields minimally coupled to higher derivative gravity. We consider the extension of Einstein-Hilbert action with $R^2$ corrections Eq.~(\ref{action}) or corrections with non-local terms Eq.~(\ref{actionnon}). In fact, in former case, beside of propagation of a massive scalar and a massless tensor fields, there are also nonphysical massless spin zero ghost and the usual massive spin 2 ghost. In latter theory, the spectrum of the theory includes a massless graviton, a massless scalar and a ghostlike massive scalar.

We expand the action of fermions Eq.~(\ref{faction}) around the flat space background and obtain two fermion-one graviton vertex Eq.~(\ref{vertex}). Then, by considering the scattering amplitude of two fermions, we obtain the non-relativistic limit and compute the potential for fermion-fermion interaction which is the usual Newtonian potential corrected with Yukawa-like terms.

We also briefly discussed the astronomical effects of such Yukawa-like potential by computing the gravitational pressure of a homogenous spherical star and used it for a white dwarf where the gravitational is in equilibrium to electron-degeneracy pressure. This equalization would give us the tree level quantum correction of Chandrasekhar radius which, as one may expect, is very small.

\acknowledgements
Mohammad A. Ganjali would like to thanks Dr. Farzan Momeni for useful discussion about white dwarf and Chandrasekhar limit.
Mohammad A. Ganjali would also like to thanks the Kharazmi university
for supporting the paper with grant.

%%%%%%%%%%%%%%%%%%%%%%%%%%%%%%%%%%%%%%%%%%%%%%%%%%%%%%%%%%%%%%%%%%%%%%%%%%%%%%%%

\end{document}